\documentclass[]{interact}

\usepackage{epstopdf}% To incorporate .pdf illustrations using PDFLaTeX, etc.
\usepackage{subfigure}% Support for small, `sub' figures and tables
\usepackage{upgreek}
\usepackage{circuitikz}
\usepackage{siunitx}

\usepackage[numbers,sort&compress,merge]{natbib}% Citation support using natbib.sty
\bibpunct[, ]{(}{)}{,}{n}{,}{,}% Citation support using natbib.sty
\renewcommand\bibfont{\fontsize{10}{12}\selectfont}% Bibliography support using natbib.sty
\renewcommand\citenumfont[1]{\textit{#1}}% Citation numbers in italic font using natbib.sty
\renewcommand\bibnumfmt[1]{(#1)}% Parentheses enclose ref. numbers in list using natbib.sty

\theoremstyle{plain}% Theorem-like structures

\theoremstyle{definition}

\theoremstyle{remark}

\begin{document}

\articletype{}

\title{Sympathetic Cooling of Protons and Antiprotons with a Common Endcap Penning Trap}

\author{
\name{M. Bohman\textsuperscript{a,b}\thanks{CONTACT Matthew Bohman. Email: matthew.bohman@mpi-hd.mpg.de}; A. Mooser\textsuperscript{b}; G. Schneider\textsuperscript{b,c}; N. Sch{\"o}n\textsuperscript{c}; M. Wiesinger\textsuperscript{a}; J. Harrington\textsuperscript{a}; T. Higuchi\textsuperscript{b,d}; H. Nagahama\textsuperscript{b}; S. Sellner\textsuperscript{b}; C. Smorra\textsuperscript{b}; K. Blaum\textsuperscript{a}; Y. Matsuda\textsuperscript{d}; W. Quint\textsuperscript{e}; J. Walz\textsuperscript{c,f}; S. Ulmer\textsuperscript{b}}
\affil{\textsuperscript{a}Max-Planck-Institut f{\"u}r Kernphysik, Saupfercheckweg 1, D-69117 Heidelberg, Germany; \textsuperscript{b}RIKEN, Ulmer Fundamental Symmetries Laboratory, 2-1 Hirosawa, Wako, Saitama 351-0198, Japan; \textsuperscript{c}Institut f{\"u}r Physik, Johannes Gutenberg-Universit{\"a}t D-5509 Mainz, Germany; \textsuperscript{d}Graduate School of Arts and Sciences, University of Tokyo, Tokyo 153-8902, Japan, \textsuperscript{e}GSI - Helmholtzzentrum f{\"u}r Schwerionenforschung GmbH, D-64291 Darmstadt, Germany; \textsuperscript{f}Helmholtz-Institut Mainz, D-55099 Mainz, Germany}
}

\maketitle

\title{Sympathetic Cooling of Single Protons with a Common Endcap Penning Trap}

\maketitle

\begin{abstract}
We present an experiment to sympathetically cool protons and antiprotons in a Penning trap by resonantly coupling the particles to laser cooled beryllium ions using a common endcap technique. Our analysis shows that preparation of (anti)protons at mK temperatures on timescales of tens of seconds is feasible. Successful implementation of the technique will have immediate and significant impact on high-precision comparisons of the fundamental properties of protons and antiprotons. This in turn will provide some of the most stringent tests of the fundamental symmetries of the Standard Model.
\end{abstract}


\section{Introduction}
Experiments that compare the basic properties of matter/antimatter counterparts with high precision provide stringent tests of charge-parity-time (CPT) invariance.
As the fundamental symmetry of the Standard Model, any measured CPT-violation will provide a clear indication of new physics. Current experimental limits constrain CPT violating effects to a fractional precision of $\num{0.5e-12}$ for leptons  \cite{Van87, Mit99}, $\num{1.3e-18}$ for mesons \cite{Sch95} and \num{2.0e-10} for hydrogen and antihydrogen \cite{Ahm17, Par11}. Our experiments target comparisons of the fundamental properties of protons (p) and antiprotons ($\bar{\text{p}}$), chiefly by measuring charge-to-mass ratios $(q/m)_{\text{p},\bar{\text{p}}}$ and magnetic moments $\mu_{\text{p},\bar{\text{p}}}$ in Penning traps. To determine the antiproton-to-proton charge-to-mass ratio we compare cyclotron frequencies $\nu_{\textrm{c}}=(q\cdot B_0)/(2\pi\cdot m)$ of negative hydrogen ions and antiprotons in the magnetic field $B_0$ of the Penning trap. Based on such measurements $(q/m)_{\bar{\text{p}}}/(q/m)_{\text{p}}$ was determined with an experimental uncertainty of $\num{69e-12}$ \cite{Ulm15}. Magnetic moment measurements, however, rely on the determination of $\nu_{\textrm{c}}$ and the spin precession frequency $\nu_{\textrm{L}}=(\mu_{\text{p},\bar{\text{p}}}/\mu_{\text{N}})\cdot \nu_{\textrm{c}}$, where $\mu_{\text{N}}$ is the nuclear magneton. The determination of $\nu_{\text{L}}$ utilizes the continuous Stern-Gerlach effect \cite{Deh86} for non-destructive detection of the particle's spin eigenstate. This method has been applied successfully in measurements of the magnetic moments $\mu_{\text{e}^-,\text{e}^+}$ of the electron and the positron \cite{Van87}, and later, in determinations of the $g$-factor of the bound electron \cite{Ver03, Stu14}. However, the mass of the proton reduces the magnetic moment by a factor of 660, making the application of the continuous Stern-Gerlach effect much more difficult.\\

Within the physics program of our collaboration we have performed proton and antiproton magnetic moment measurements with fractional uncertainties on the parts per billion \cite{Moo14} and parts per million level \cite{Nag17}, respectively. These experiments are limited by the energy in the harmonic modes of the particle in the trap and considerably benefit from the development of experimental techniques that provide a fast and deterministic way to reduce the energy of the particle to $E/k_B < \SI{100}{\milli\kelvin}$. To overcome this inevitable experimental reality, we have developed a new experiment that will sympathetically cool protons and antiprotons by resonantly coupling the particles to laser-cooled beryllium ions with a common endcap technique, developing on an early idea by Heinzen and Wineland \cite{Win90}. \\

In this manuscript we first provide a detailed discussion of the continuous Stern-Gerlach effect in proton and antiproton magnetic moment measurements and the limitations we now face. We summarize results of our feasibility studies to apply sympathetic cooling to single particles by using a \emph{common endcap} technique. Based on our analysis we anticipate preparation of protons/antiprotons at temperatures below 30 mK, with a timescale of tens of seconds. Finally, we describe the new apparatus we are currently commissioning in the context of an improved measurement of the proton magnetic moment.\\

\subsection{Continuous Stern-Gerlach Effect}
A Penning trap is formed by the superposition of a homogeneous magnetic field ${\mathbf{B} = B_{0} \hat{\mathbf{z}}}$ and a static electric quadrupole potential. The motion of a charged particle in such a trap is composed of three independent harmonic modes with eigenfrequencies $\nu_z$ along the magnetic field lines, as well as $\nu_+$, and $\nu_-$ perpendicular to $\mathbf{B}$. The mode energies $E_{+} = h \nu_{+}(n + \frac{1}{2})$, $E_{z} = h \nu_{z}(k + \frac{1}{2})$, $|E_{-}| = h \nu_{-}(l + \frac{1}{2})$ where $h$ is the Planck constant, and $n$, $k$, $l$ are the principal quantum numbers, are referred to as the modified cyclotron, axial, and magnetron modes. The corresponding frequencies are measured using an image current detection technique \cite{Fen96, Deh86, Nag16} and the cyclotron frequency is determined through the Brown-Gabrielse invariance theorem $ \nu_{\textrm{c}}^{2} = \nu_{+}^2 + \nu_{z}^2 + \nu_{-}^2$ \cite{Bro87}. In addition the spin of the particle splits the energy of the states by $E_{m_s} = |E_{+}(m_s=+\frac{1}{2}) - E_{+}(m_s=-\frac{1}{2})| = h \nu_{\textrm{L}}$. However, while the frequency $\nu_{\text{L}}$ must be determined in magnetic moment measurements, it is not accompanied by a detectable shift of charge, and cannot be directly measured by image current detection. Instead, we utilize the continuous Stern-Gerlach effect by superimposing a magnetic bottle, $B(z)=B_0+B_2z^2$ with strength $B_2$ to couple the total magnetic moment of the particle in the trap, $\mu_{\text{tot}}$ to the axial frequency $\nu_z$. Here $\mu_{\text{tot}}= \mu_{+}+\mu_{-}+\mu_{\text{p},\bar{\text{p}}}$ is the sum of the magnetic moments generated by the modified cyclotron and magnetron motion $\mu_{+}$ and $\mu_{-}$, respectively, as well as the intrinsic magnetic moment of the particle $\mu_{\text{p},\bar{\text{p}}}$. The shift of the axial frequency in the magnetic bottle is a function of the principal quantum numbers $n$ and $l$, along with the spin state $m_s$ and is given by:

\begin{equation}
    \Delta\nu_z(n,l,m_s)=\frac{h \nu_+}{4\pi^2 m_{p, \bar{p}}\nu_z}\frac{B_2}{B_0}\left[\left(n+\frac{1}{2}\right)+\frac{\nu_-}{\nu_+}\left(l+\frac{1}{2}\right)+g_{p, \bar{p}}m_s\right]~.
\label{eq:Bottleshift}
\end{equation}

The coupling introduced by the magnetic bottle enables the measurement of the spin transition rate as a function of an applied rf drive by providing a nondestructive determination of the particle's spin eigenstate. From a fit of the well understood resonance line \cite{Bro84, Bro85} to the measured data, $\nu_{\text{L}}$ can be extracted. However, compared to the electron, the mass of the proton reduces $\mu_{p,\bar{p}}$ and makes the application of this technique much more difficult.
Therefore we superimpose a magnetic bottle with a strength of $B_2\approx \SI[per-mode = fraction]{300000}{\tesla\per\meter\tothe{2}}$ on our trap, which is about 2000 and 30 times larger than in \cite{Van87} and \cite{Stu14}, respectively. In this strong inhomogeneity a spin transition causes an axial frequency shift of about $\SI{200}{\milli\hertz}$ at $\nu_z \approx \SI{650}{\kilo\hertz}$: the signal needed to determine $\nu_{\text{L}}$ and thus the $g$-factor, $g_{p,\bar{p}}$, the magnetic moment $\mu_{p,\bar{p}}$ in units of the nuclear magneton $\mu_N$. Our group has managed to apply this method to observe proton \cite{Ulm11, Moo13} and antiproton \cite{Smo17} spin transitions by cooling the modified cyclotron mode of the particle to below $E_+/k_B<\SI{0.6}{\kelvin}$ \cite{Moo13}. Only at such low temperatures is the axial frequency $\nu_z$ stable enough to observe the frequency shifts $\Delta\nu_{z,\text{SF}}(\Delta m_s=1)$, caused by a spin flip. Our observation that the axial frequency stability $\Xi_z(E_+)$ is a function of the cyclotron energy can be understood by considering transition rates in the cyclotron mode, which are driven by

\begin{figure}
\centering
\subfigure[]{
\resizebox*{6.5cm}{!}{\includegraphics{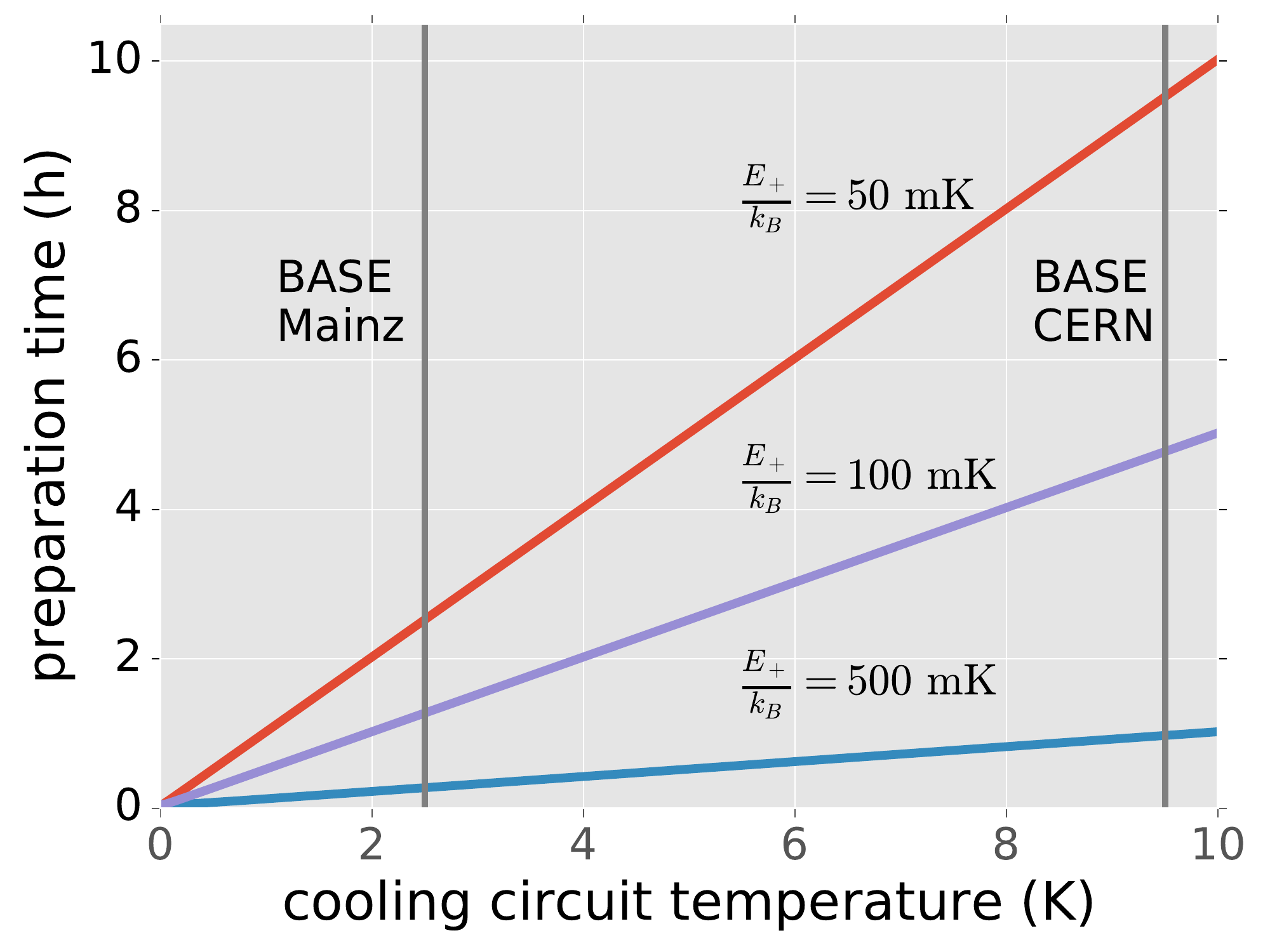}}}\hspace{5pt}
\subfigure[]{
\resizebox*{6.5cm}{!}{\includegraphics{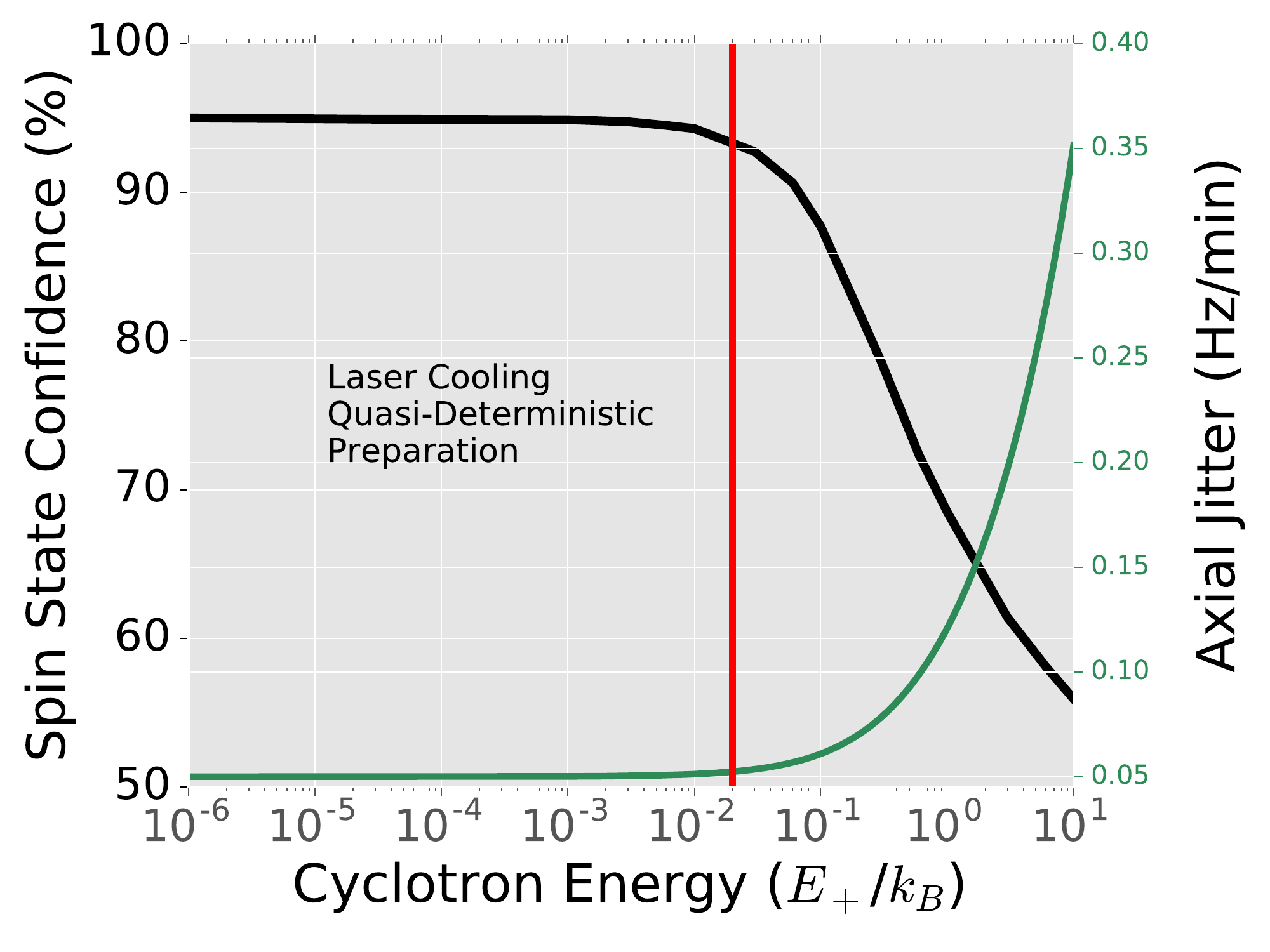}}}
\caption{(a) Approximate preparation times required to prepare an (anti-)proton at a cyclotron temperature of $E_+/k_B=0.05\,$K (blue), $0.1\,$K (green), and $0.5\,$K (orange). The vertical grey lines represent the different conditions for the BASE CERN antiproton experiment and the BASE Mainz proton experiment \cite{Smo15}. (b) Illustration of the spin state detection fidelity and the axial frequency fluctuation as a function of the cyclotron temperature $E_+/k_B=\SI{0.5}{\kelvin}$. At radial temperatures below 0.1$\,$K we achieve a spin state identification fidelity close to 100$\,\%$. At present, we use particles with $E_+/k_B<0.6\,$K, prepared by resistively coupling to a detector at $\approx 2\,$K (Mainz) and $\approx \SI{9.5}{\kelvin}$ (CERN).}
\label{Cyclotron Fluctuations}
\end{figure}

electric field noise with a power spectrum density $\langle e_n(t),e_n(t-\tau)\rangle$ \cite{Sav97}. In a given cyclotron quantum state $n$ this noise induces transitions at a rate of
\begin{eqnarray}
    \frac{dn}{dt}\approx\frac{q^2}{2 m_{p, \bar{p}} h \nu_{+}}n \Lambda^2 \langle e(t),e(t-\tau)\rangle\,,
\end{eqnarray}
where $\Lambda$ is a parameter that defines a length scale of the electrical field and the scaling $\propto n$ arises from the harmonic oscillator transition matrix elements. Note that for typical experimental parameters such as in \cite{Moo14} as well as the experiment described in the following sections, a single cyclotron quantum jump in the magnetic bottle already induces an axial frequency shift of $\SI{70}{\milli\hertz}$. Thus, depending on the experiment, in an alternating sequence of axial frequency measurements and spin transition drives, typically between \SIlist{90;180}{\second} each, not more than two $n$-transitions can be accepted, as the information needed to determine the spin state becomes inseparable from the additional noise in the axial frequency.\\

Up to now we have used selective resistive cooling techniques to prepare particles below a threshold $E_+/k_B<\SI{1}{\kelvin}$, by coupling to resonant cooling circuits at effective temperatures which are on the order of the physical temperature of our experiments. With our current experimental parameters, the time required to prepare a particle at a certain temperature is shown in Fig$.\,$\ref{Cyclotron Fluctuations}(a). Together with the spin state preparation described in \cite{Moo13, Smo17} this procedure consumes about 50$\,\%$ to 60$\,\%$ of our total measurement time budget. As summarized in Tab$.\,$1, sampling a full $g$-factor resonance requires about two months of continuous measurements, with at least one month entirely due to particle cooling and the ambiguous readout of the spin state due to thermally induced frequency fluctuations. The successful application of sympathetic cooling of protons and antiprotons to deterministically low temperatures by coupling the particles to laser-cooled ions will drastically reduce the cycle time of our experiments, as illustrated in Fig$.\,$ \ref{Cyclotron Fluctuations}(b). This will enable measurements at higher statistics and improved time resolution, as well as precision studies at lower statistical uncertainty, and time resolved measurements. For instance, sidereal variations of fundamental properties of the particle may be induced by parameters coupling to Lorentz-violating cosmic background fields \cite{Kos11} could be further constrained simply by increasing the number of data points in a given time window. Moreover, the reduced cyclotron energy $E_+$ creates a path toward resolving coherent spin-dynamics, reducing the widths of $g$-factor resonance lines and facilitating the application of phase sensitive detection techniques in direct measurements of the cyclotron frequency \cite{Stu11}. The latter will have impact on both, high-precision measurements of the proton and antiproton magnetic moments as well as comparisons of the proton-to-antiproton charge-to-mass ratio.
\begin{table}
\tbl{Time budget for the most recent proton \textit{g}-factor measurement. A typical measurement campaign builds up a \textit{g}-factor resonance with around 1000 data points.}
{\begin{tabular}{lcccccc} \toprule
Measurement Type & Mean (min) & Median (min) & Mean Fractional Time (\%) & Median Fractional Time (\%)\\ \midrule
Spin State Detection & 45 & 40 & 43 & 43\\
Precision Frequency Measurement & 32 & 32 & 31 & 34 \\
Cold Particle Search & 27 & 22 & 26 & 23\\
Total & 104 & 94 & 100 & 100\\ \bottomrule
\end{tabular}}
\label{sample-table}
\end{table}

\subsection{Sympathetic Cooling}
To achieve lower particle temperatures, we can take advantage of the major successes in cold ion traps experiments over the past few decades. Doppler cooled ions, for example, at sub mK temperatures, are the foundation of successful trapped ion quantum information experiments \cite{Mon16, Win95, Bla16}. Likewise ions in rf traps \cite{Mon95}, and more recently Penning traps \cite{Goo16}, are routinely prepared in the ground state. Similarly, sympathetic cooling, in which a laser cooled ion couples via the Coulomb interaction to another ion without a suitable cooling transition, has been demonstrated for a variety of ions \cite{Lar86,Bli02, Bar03} and implemented for modern metrology and spectroscopy experiments \cite{Sch15}.\\

Our new experiment utilizes this existing technology and employs a novel cooling scheme first proposed in \cite{Win90} in which a stack of two compensated, cylindrical Penning traps are coupled by a common endcap as shown in Figure 2(a). In this scheme, a cloud of beryllium ions with axial temperatures near the Doppler limit of around \SI{0.5}{\milli\kelvin} is brought into resonance with a single proton by tuning the trap potentials. Once on resonance, the particle thermalizes via image currents induced in the common endcap, and the axial temperature of the proton will be limited only by the coupling strength of the interaction compared to other sources of heating. Finally, the cyclotron energy can be reduced by applying an rf pulse to couple the axial energy to the cyclotron energy so that the resulting mode energy is given by frequency ratio $\frac{E_{z}}{E_{+}} = \frac{\omega_{z}}{\omega_{+}}$. \\

The following sections provide a theoretical basis for the exchange of energy between a single proton and Doppler cooled beryllium ions across a common Penning trap endcap. We discuss various sources of decoherence and show that a realistic experiment can be constructed so that the coupling strength of the cooling interactions are the dominant sources of energy exchange in the system. We also show the details of a new experiment that demonstrates these techniques in a measurement of the proton $g$-factor.

\section{Common Endcap Coupling}

\subsection{Coupled Harmonic Oscillators}
In the most basic implementation, sympathetic cooling can be achieved simply by storing a laser cooled ion in the same potential well as the particle to be sympathetically cooled with the resulting system being nothing more than two coupled quantum oscillators. However, this approach fails almost entirely when applied to negatively charged particles, such as the antiproton, due to the lack of Coulomb repulsion. Despite some early successes in coupling oscillators in different traps \cite{Bro11, Har11}, such an approach remains difficult and hard to translate to high precision experiments in Penning traps. As a result, we aim for a common endcap approach, in which an electrode shared by two traps mediates the Coulcomb exchange between the laser cooled ion and the particle of interest across a macroscopic distance and the traps themselves provide the necessary repulsive force.\\

To quantitatively describe this coupling it is useful now to switch to a circuit model with identical dynamics. A particle, or cloud of $N_{i}$ particles, with mass $m_i$ and charge $q_i$ in a Penning trap can be described by a resonant RLC circuit \cite{Fen96, Win75} with equivalent inductance $L_{i}$, capacitance $C_{i}$, and resistance $R_i$

\begin{equation}
 L_{i} = \frac{4m_{i}d_{i}^{2}}{N_{i}q^{2}}
\end{equation}
\begin{equation}
 C_{i} = \frac{1}{\omega_{z}^2 L_{i}} = \frac{q^{2}}{4N_{i}\omega_{z} m_{i}d_{i}^{2}}
\end{equation}
\begin{equation}
 R_{i} = \gamma L_{i} = \frac{\gamma4m_{i}d_{i}^{2}}{N_{i}q^{2}}
\end{equation}
where $d_{i}$ is a parameter that depends on trap geometry \cite{Bro87}, and $\gamma$ is the damping of the equivalent circuit. Following \cite{Win90}, the common endcap coupling can be represented simply by adding a capacitor, $C_{T}$ in parallel to the equivalent circuits of the two traps as shown in Figure 2. The solutions to the Hamiltonian of this system \cite{Win90}  describe an oscillatory exchange of energy between the two ion-trap systems, just as in the direct Coulomb coupling scheme, and are given by:

\begin{equation}
    \omega_{\pm}^2 = \frac{1}{2}[(\omega_{1}^{'})^2 + (\omega_{2}^{'})^2] \pm  \frac{1}{2}\sqrt{[(\omega_{1}^{'})^2 + (\omega_{2}^{'})^2]^2 + 4g^4}       
\end{equation}

where $g^2 = \left(L_{1}L_{2}C_{T}^2\right)^{-1/2}$ and $\omega_{i}^{'}$ is the reduced resonance frequency $\omega_{i}^{'} = \left(L_{i}C_{i}^{'}\right)^{-1/2}$ with $C_{i}^{'} = \frac{C_{i} C_{T}}{C_{i} + C_{T}}$. In the resonance case, i.e. $\bar{\omega} = \frac{1}{2}[(\omega_{+}) + (\omega_{-})] \approx (\omega_{1}^{'}) \approx (\omega_{2}^{'})$ the energy exchange is characterized by $\omega_{ex} = \frac{1}{2} (\omega_{+} - \omega_{-})$, giving rise to an exchange time: 

\begin{equation}
        \tau_{ex} = \frac{\pi}{2\omega_{ex}} = \pi \bar{\omega}\sqrt{L_1 L_2} C_T.
\end{equation}

If we now consider $\bar{\omega}$ to be the shared axial frequency of two coupled ions or ion clouds, Equation (7) becomes

\begin{equation}
        \tau_{ex} = \frac{\pi\omega_{z}{d_{1}d_2}}{\sqrt{N_1}\sqrt{N_2}}C_{T}\frac{\sqrt{m_{1}m_{2}}}{|q_1q_2|},
\end{equation}
in which $\omega_{z}$ is the axial frequency of the particles in both traps. In fact, more generally, the time constant 
\begin{equation}
    \tau = \alpha \omega_s \frac{\sqrt{m_1m_2}}{|q_1q_2|},
\end{equation}
describes the general resonant coupling of two oscillators with shared frequency $\omega_s$, where $\alpha$ is a proportionality factor that determines the strength of the coupling and depends on the medium through which energy is exchanged.\\

We can now insert experimental parameters necessary to sympathetically cool a proton in this scheme. In a realistic proton \textit{g}-factor experiment, $\omega_z/2\pi$ is on the order of a few hundred kHz, determined by the potentials applied to the trap electrodes. In our planned proton $g$-factor experiment the central potential is approximately 1 V, with $\omega_{z} \approx 2\pi \times \SI{500}{\kilo\hertz}$. Meanwhile, in small but machinable, cylindrical traps, $d_i$ is typically a few mm, while $m_2$ and $q_2$ are set by the choice of ion species to laser cool. The lightest ion with a suitable cooling transition is beryllium, which conveniently is one of only a few species that can be cooled by a single laser. There is then a clear incentive to maximize the number of particles and to minimize the trap capacitance $C_T$. In an experiment with a single trapped proton, a cloud of beryllium ions, and a symmetric trap design Equation (8) becomes
\begin{equation}
        \tau_{ex} = \frac{\pi\omega_{z}d^2}{\sqrt{N_{Be}}} C_T  \frac{\sqrt{m_p m_{Be}}}{e^2}.
\end{equation}
After designing and simulating a trap with minimal capacitance, we can achieve $C_T \approx \SI{5}{\pico\farad}$ and $d = \SI{4.6}{\milli\metre}$ with an additional capacitance of around \SI{5}{\pico\farad} due to the wiring. Assuming the experiment is normally operated with 100 beryllium ions we obtain a coupling time of $\tau_{ex} \approx \SI{55}{\second}$.

\begin{figure}
\centering
\subfigure[]{
\resizebox*{5cm}{!}{\includegraphics{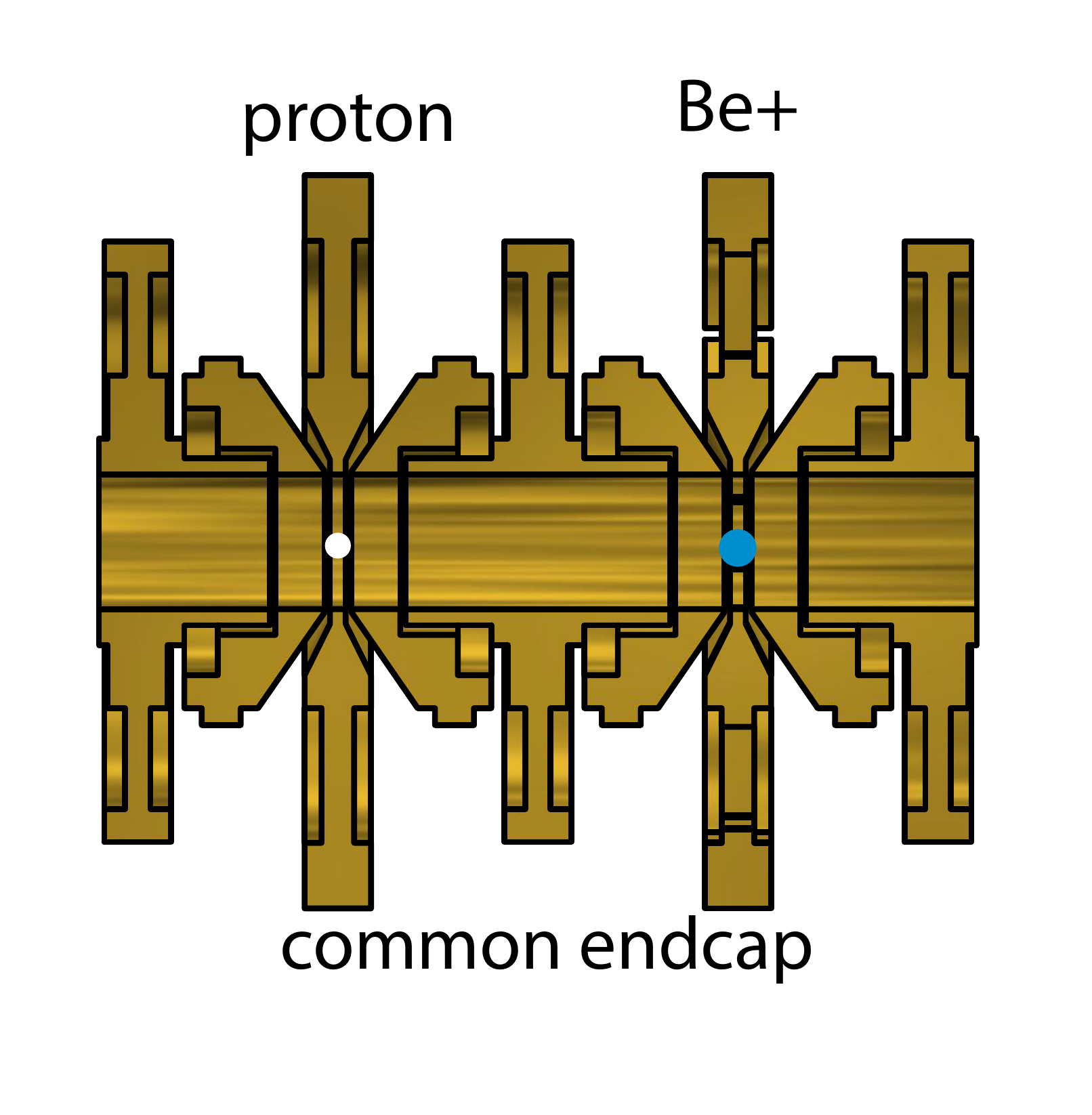}}}
\subfigure[]{
\resizebox*{5cm}{!}{\includegraphics{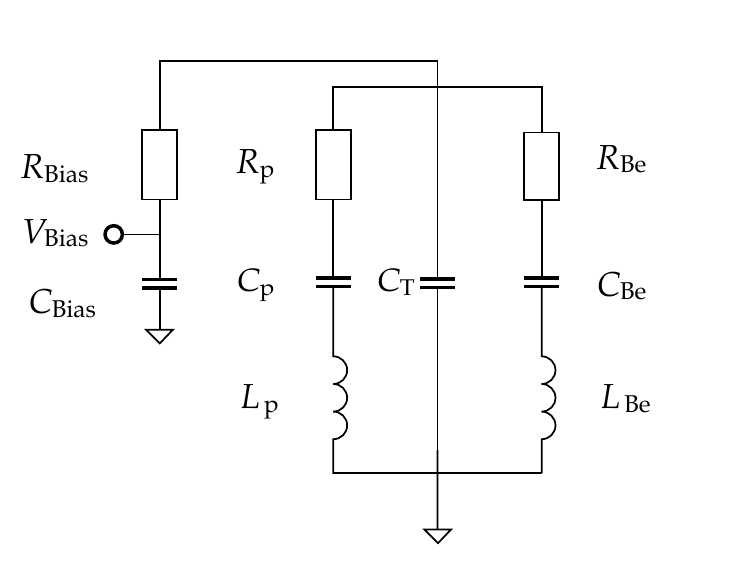}}}
\caption{(a) Penning traps for a proton and beryllium ion cloud connected by a common endcap. Each trap has an outer endcap electrode, two inner correction electrodes, a central ring electrode and one shared endcap. The unique shape of the trap electrodes has been chosen to minimize the capacitance $C_T$ of the common endcap to ground and thus minimize the exchange time. (b) Circuit representation of the trap configuration on the left. The series RLC circuits to the left and right of the common endcap, $C_T$, are the equivalent circuits of the proton-trap and ion-trap systems, respectively.}
\label{Coupling}
\end{figure}

\subsection{Sources of Heating}
While a promising first step, this analysis is complicated by the fact that there are additional sources of heating that must interact with the proton on a time scale much slower than that of the coupling. Two critical sources are: noise to the biasing of the common endcap and off-resonant interactions with a detection system in the two traps. Following \cite{Dan09}, in which a similar coupling scheme connects two ions by a macroscopic wire, the interaction timescale of heating due to a resistor with Johnson noise $ P_{noise} = k_B T \Delta \nu$, in a bandwidth $\Delta \nu$, is given by
\begin{equation}
        \tau_{noise} = \frac{h}{k_B T R} \sqrt{\frac{L_{i}}{C_{i}}}
\end{equation}
In \cite{Dan09}, this problem is solved by simply floating the coupling wire but this presents serious obstacles as the wire or electrode is then allowed to freely accumulate charge. Our solution, shown in Figure 2(b), is to block the high frequency image currents with a resistor connected to the bias supply, while a capacitance to ground introduces a decoupling that reduces the effective resistance of the biasing network to $R_{eff} = \mathrm{Re}(Z_R + Z_C)$ where $Z_R$ and $Z_C$ are the impedances of the resistor and capacitor, ultimately yields a coupling time constant of, $\tau_{\text{p-bias}} \approx \SI{e5}{\second}$ for the proton and $\tau_{\text{Be-bias}} \approx \SI{e4}{\second}$ for the beryllium ions.\\

A more serious problem comes from the necessity of measuring the axial frequencies of the two ion species to bring them into resonance in the two traps. Typically, this is achieved by connecting a tuned RLC circuit, acting as a thermal bath, in parallel to the trap as described in \cite{Win75, Fen96, Nag16}. This circuit can be shown to interact with the proton with a time constant given by: 

\begin{equation}
    \tau_{D} = 1/\gamma_{D} = \left(\frac{R_{eff}q_{i}^2N_{i}}{m_{i}d_{i}}\right)      
\end{equation}

where $\gamma_{D}$ is the damping of the equivalent circuit with $R_{eff} = \textrm{Re}(Z)$, where $Z$ is the impedance of the circuit. However, with no detuning, the resonant coupling $\tau_{D}$ is on the order of $\SI{100}{\milli\second}$ - completely overwhelming the coupling of the beryllium ions to the proton. As seen in Figure 3, a switchable parallel capacitance shifts the resonance by $\sim$ \SI{200}{\kilo\hertz}, resulting in an off-resonant coupling time of $\tau_{D} \approx \SI{e6}{\second}$.

\begin{figure}
\centering
{\resizebox*{10cm}{!}{\includegraphics{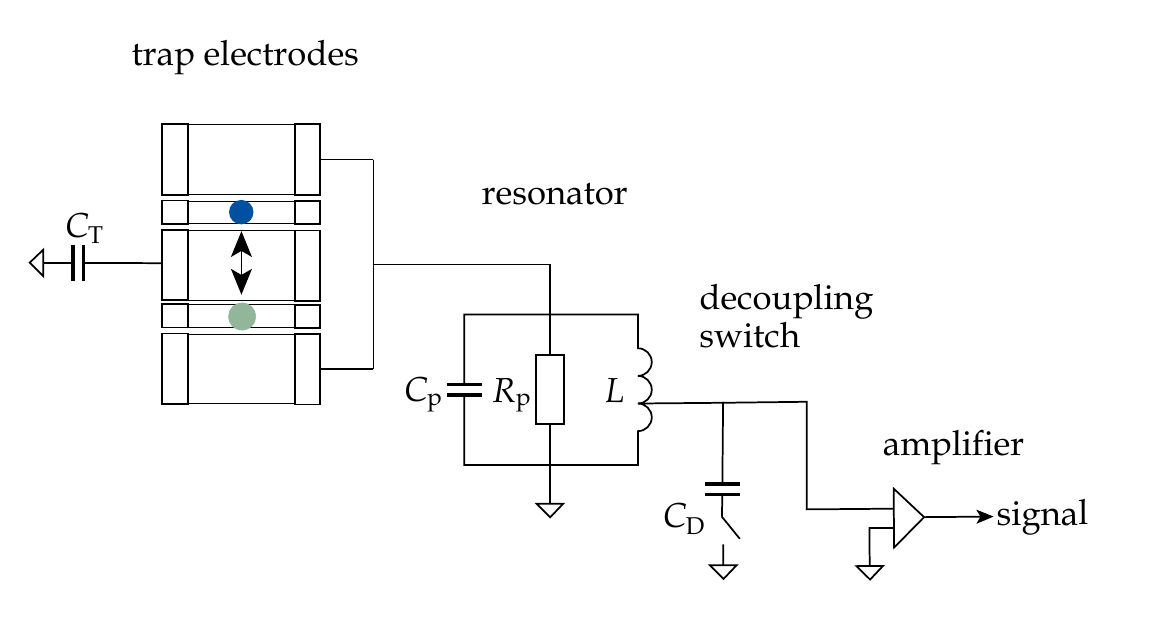}}}\hspace{50pt}
\caption{The common endcap technique and detection system. A tuned resonator connected to the traps storing the proton and beryllium ions allows the detection of the axial frequencies while a switched capacitor, $C_D$ shifts the resonator to allow particle-ion interactions to dominate.} \label{Decoupled Traps}
\end{figure}

\subsection{Expected Results}
With these results, we can now consider the resulting temperatures of the common endcap system shown in Figure 4. Extending the argument of \cite{Win75}, we can model the energy exchange in the system by considering the coupling of the laser, detector, and biasing network at temperatures $T_L$, $T_{d}$, and $T_{bias}$, respectively, to the proton and beryllium ions with interaction times $\tau_{i-j}$, in which the indices $i-j$ indicate the interaction of the $i$th component of the system with the $j$th component of the system. In the steady state, we find:

\begin{equation}
    \begin{split}
         0 = - k_{B}\left(\frac{1}{\tau_{\text{D-p}}} + \frac{1}{\tau_{\text{Be-p}}} + \frac{1}{\tau_{bias}}\right)T_p
         + \frac{k_B}{\tau_{\text{D-p}}} T_{D} + \frac{k_B}{\tau_{\text{Be-p}}} T_{Be} + \frac{k_B}{\tau_{bias}} T_{bias}
    \end{split}
\end{equation}

\begin{equation}
        0 = - k_{B}\left(\frac{1}{\tau_{Be–p}} + \frac{1}{\tau_{D‒Be}} + \frac{1}{\tau_{L}} + \frac{1}{\tau_{bias}}\right) + \frac{k_B}{\tau_{D‒Be}} T_{D} + \frac{k_B}{\tau_\text{Be-p}} T_{p} + \frac{k_B}{\tau_{L}} T_{Be} + \frac{k_B}{\tau_{bias}} T_{bias}
\end{equation}

\begin{figure}
\centering
{\resizebox*{10cm}{!}{\includegraphics{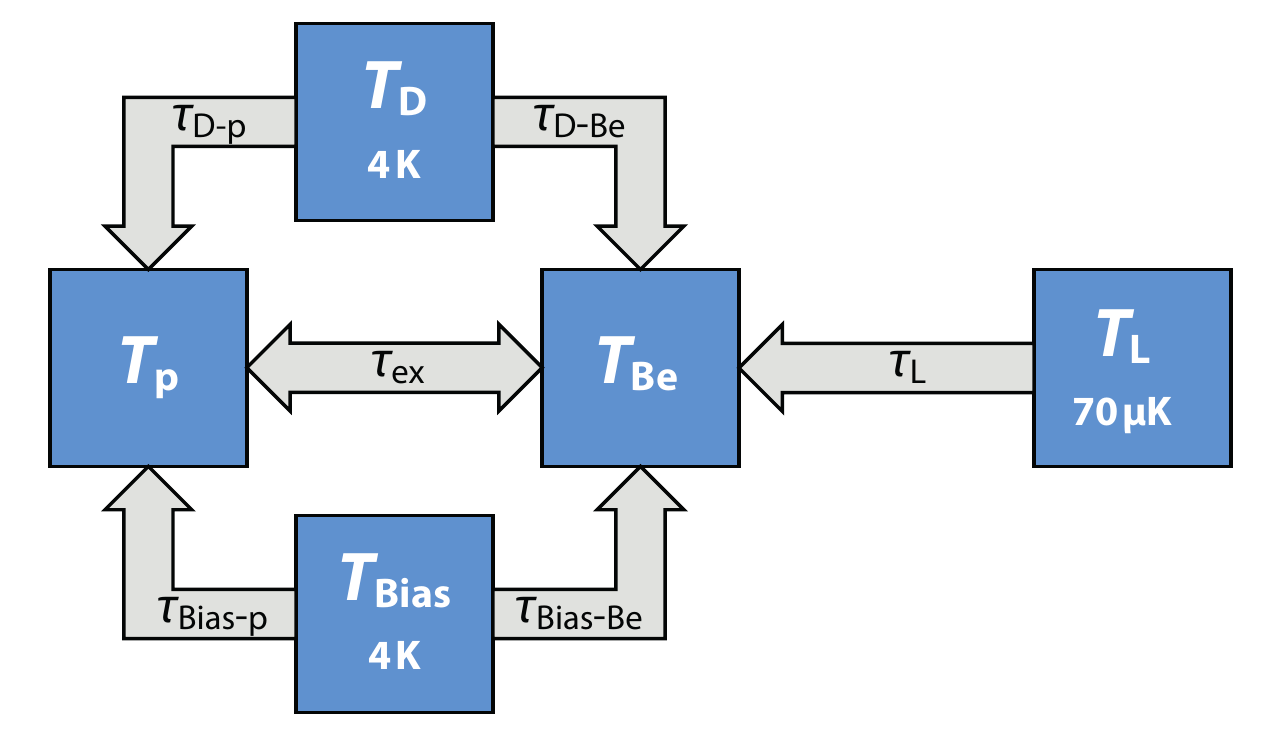}}}\hspace{40pt}
\caption{The detector circuit and biasing network are modelled as thermal baths at $T_D = \SI{4}{\kelvin}$ that interact with the proton and beryllium ion cloud with characteristic time scales $\tau_{D-p}$,$\tau_{\text{D-Be}}$, $\tau_{\text{Bias-p}}$, and $\tau_{\text{Bias-Be}}$, respectively. Similarly, the laser is modelled as a thermal bath at the Doppler limit $T_L \approx \SI{0.5}{\milli\kelvin}$ that interacts with the beryllium ions alone with a characteristic time scale $\tau_L$. Finally the common endcap interaction is given by $\tau_{ex}$ so that we obtain effective steady state temperatures of the proton and the beryllium ion cloud $T_p$ and $T_{Be}$.} \label{Interaction Diagram}
\end{figure}

which produces an algebraic solution for $T_p$ and $T_{Be}$. Using experimentally achievable values of $\tau_{\text{Be-p}} \approx \SI{55}{\second}$, $\tau_{\text{D-p}} \approx \SI{2e6}{\second}$, $\tau_{{D}-{Be}} \approx \SI{2e5}{\second}$, and approximating $\tau_{L} \approx \frac{1}{\Gamma_{ed}}$, where $\Gamma_{ed}$ is the linewidth of the cooling transition $\gamma/2\pi \approx \SI{19.4}{\mega\hertz}$, we can find effective steady state temperatures for the proton and the beryllium ion cloud, $T_{p}$, $T_{Be}$ given a detector temperature $T_D = \SI{4}{\kelvin}$ and an effective `laser' temperature near the Doppler limit $T_L \approx \SI{0.5}{\milli\kelvin}$. Ultimately, these values produce effective axial temperatures around a few mK and, after coupling the axial mode to the cyclotron mode by applying an rf pulse at a sideband frequency, we find cyclotron temperatures of $T_c \approx T_z \frac{\nu_{\textrm{c}}}{\nu_z} < \SI{30}{\milli\kelvin}$.

\section{Experimental Realization}
The experiment we have designed to realize this sympathetic cooling scheme is a heavily modified version of the double trap apparatus used in \cite{Moo14}. It consists of five, cylindrical, compensated Penning traps linearly connected in the bore of a 1.95 T, superconducting magnet as shown in Figure 5. From left to right, these are the source trap (ST), analysis trap (AT), precision trap (PT), coupling trap (CT), and beryllium trap (BT). The AT and the PT together form the `double trap' \cite{Moo14} in which the PT with high magnetic field homogeneity is used for high precision frequency measurements and the AT with a strong, superimposed magnetic bottle is used for spin state analysis. The new ST is used to prepare particles, while the BT stores laser cooled beryllium ions in a rotating wall potential and the CT is connected adjacently to implement the common endcap coupling. Finally, a small section to the right holds a beryllium target and optics for a pulsed Nd:YAG laser, forming an in situ ion production region.\\

\begin{figure}
\centering
{\resizebox*{15cm}{!}{\includegraphics{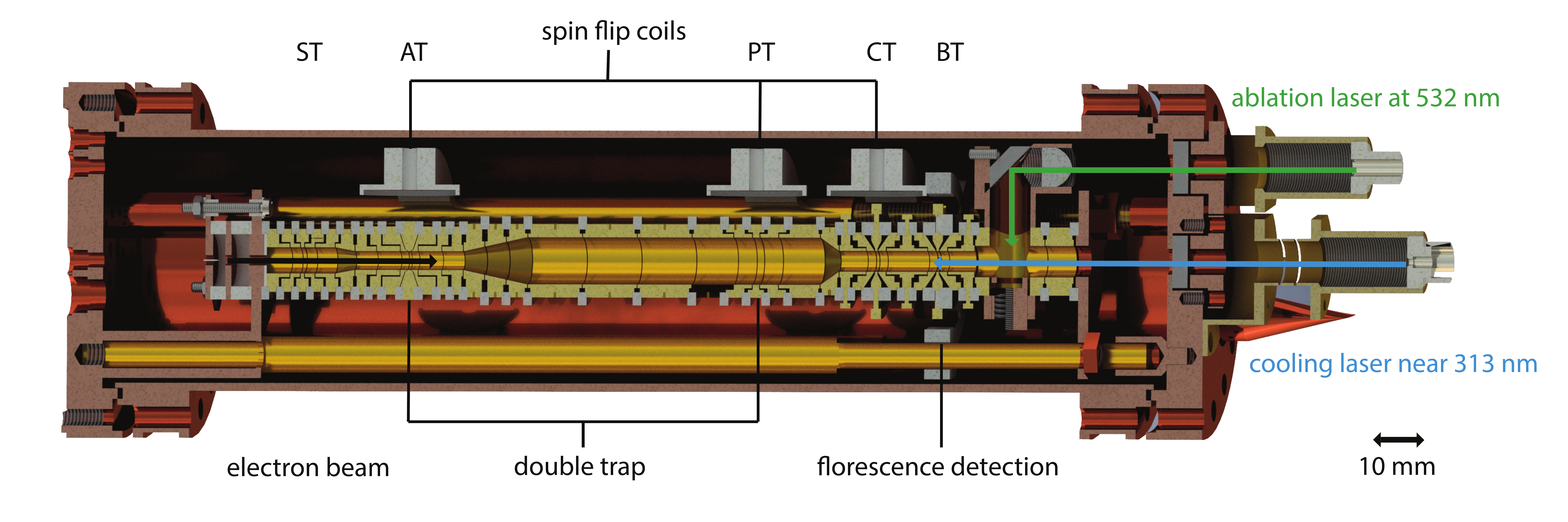}}}\hspace{50pt}
\caption{The new apparatus for an improved proton $g$-factor measurement. The five traps are labelled from right to left, as the source trap (ST), analysis trap (AT), precision trap (PT), coupling trap (CT), and beryllium trap (BT).The gold plated electrodes, in the bore of a 1.95 T superconducting magnet form different Penning traps that, in the case of the AT and PT implement the `double trap' technique used in \cite{Moo14} and in the case of the CT and BT, the common endcap cooling technique. On the right side of the apparatus, fibres couple in a pulsed, frequency doubled Nd:YAG laser for ion loading and a frequency quadrupled diode laser at \SI{313.13}{\nano\meter} for Doppler cooling \cite{Col14}. Finally, a sixfold split electrode in the BT serves as a dual purpose as a rotating wall and a florescence collection system. See text for additional details.} \label{Annotated Trap Stack}
\end{figure}

During a \textit{g}-factor measurement, beryllium atoms are first sputtered off the target and ionized by an axial electron beam. Once shuttled into the BT, a single laser along the magnetic field lines, red detuned from the $2\textrm{S}_{\frac{1}{2}}\to 2\textrm{P}_{\frac{3}{2}}$ transition near \SI{313}{\nano\meter} cools the ions and a sixfold split rotating wall electrode thermally couples the axial and radial modes of the ions. The split electrode serves an additional purpose by allowing florescence to be detected by silicon photomultipliers (SiPMs) through small sapphire blocks. On the opposite side of the apparatus, the electron beam produces protons from a target and a single proton can be prepared using standard techniques \cite{Smo15}.\\

During the coupling, the proton and the beryllium ion are brought to the CT and BT, respectively, and are each coupled to an axial detector. The potentials are then tuned so that the axial frequencies match and the detector is switched off resonance. Once the proton-ion system thermalizes the cold axial mode can be coupled to the cyclotron mode, and the particle can be transported throughout the apparatus with minimal heating. In contrast to previous measurements, the spin state can be determined immediately with extremely high fidelity, given a cyclotron temperature of even \SI{100}{\milli\kelvin}, reducing the time of the entire cooling cycle to around one minute.

\section{Outlook}
We have presented a novel experiment to perform a direct CPT test at improved precision with measurements on sympathetically cooled protons and antiprotons. To this end, the common endcap approach we have chosen to use for an upcoming proton $g$-factor measurement can be immediately applied to the antiproton. Meanwhile, a different approach pursued by a related group uses the techniques of quantum logic spectroscopy to perform an even more ambitious measurement of the proton $g$-factor with direct Coulomb coupling in a microfabricated Penning trap \cite{nie14}. However, we anticipate that the development of the common endcap technique will have many additional applications in ion trap physics. For example, the measurement of heavier nuclear magnetic moments can be measured identically to the proton magnetic moment. Similarly, the coupling and cooling scheme shown here could be used with other systems such as molecular ions, highly charged ions (HCIs), or more complex antimatter systems. In fact, since the coupling time is inversely proportional to the charge of the coupled particle, HCI experiments would benefit dramatically and can perhaps even realize coherent exchanges. More generally, the techniques described here are an additional step toward the demonstration of interacting quantum systems through macroscopic classical intermediaries. In the near future, the measurement of the proton magnetic moment, followed soon after by a similar measurement of the magnetic moment of the antiproton, will provide more than an order of magnitude improvement on CPT constraints and the most precise direct CPT comparison of single baryons.

\section*{Funding}

We acknowledge financial support by the Helmholtz-Gemeinschaft, by RIKEN Initiative Research Unit Program, RIKEN Pioneering Project Funding, RIKEN FPR Funding, the RIKEN JRA Program, the Max-Planck Society and by the EU (Marie Sklodowska-Curie grant agreement No 721559).

\bibliography{ref.bib}

\begin{thebibliography}{10}

\bibitem{Van87}
R.~S. Van~Dyck, P.~B. Schwinberg, and H.~G. Dehmelt.
\newblock New high-precision comparison of electron and positron g factors.
\newblock {\em Phys. Rev. Lett.}, 59:26--29, Jul 1987.

\bibitem{Mit99}
R.~K. Mittleman, I.~I. Ioannou, H.~G. Dehmelt, and N.~Russell.
\newblock Bound on $\mathit{CPT}$ and lorentz symmetry with a trapped electron.
\newblock {\em Phys. Rev. Lett.}, 83:2116--2119, Sep 1999.

\bibitem{Sch95}
B.~Schwingenheuer, R.~A. Briere, A.~R. Barker, E.~Cheu, L.~K. Gibbons, D.~A.
  Harris, G.~Makoff, K.~S. McFarland, A.~Roodman, Y.~W. Wah, and et~al.
\newblock $\mathit{CPT}$.
\newblock {\em Phys. Rev. Lett.}, 74:4376--4379, May 1995.

\bibitem{Ahm17}
M.~Ahmadi, B.~X.~R. Alves, C.~J. Baker, W.~Bertsche, E.~Butler, A.~Capra,
  C.~Carruth, C.~L. Cesar, M.~Charlton, S.~Cohen, and et~al.
\newblock Observation of the 1s--2s transition in trapped antihydrogen.
\newblock {\em Nature}, 541(7638):506--510, 01 2017.

\bibitem{Par11}
C.~G. Parthey, A.~Matveev, J.~Alnis, B.~Bernhardt, A.~Beyer, R.~Holzwarth,
  A.~Maistrou, R.~Pohl, K.~Predehl, T.~Udem, and et~al.
\newblock Improved measurement of the hydrogen $1s-2s$ transition frequency.
\newblock {\em Phys. Rev. Lett.}, 107:203001, Nov 2011.

\bibitem{Ulm15}
S.~Ulmer, C.~Smorra, A.~Mooser, K.~Franke, H.~Nagahama, G.~Schneider,
  T.~Higuchi, S.~Van~Gorp, K.~Blaum, Y.~Matsuda, W.~Quint, J.~Walz, and
  Y.~Yamazaki.
\newblock High-precision comparison of the antiproton-to-proton charge-to-mass
  ratio.
\newblock {\em Nature}, 524(7564):196--199, 08 2015.

\bibitem{Deh86}
H~Dehmelt.
\newblock Continuous stern-gerlach effect: Principle and idealized apparatus.
\newblock {\em Proc. Natl. Acad. Sci. USA}, 83:2291--2294, 1986.

\bibitem{Ver03}
J.~Verd{\'u}, T.~Beier, S.~Djekic, H.~H{\"a}ffner, H.-J. Kluge, W.~Quint,
  T.~Valenzuela, and G.~Werth.
\newblock {\em Measurement of the g Factor of the Bound Electron in
  Hydrogen-like Oxygen 16O7+}, pages 47--52.
\newblock Springer Netherlands, Dordrecht, 2003.

\bibitem{Stu14}
S.~Sturm, F.~Kohler, J.~Zatorski, A.~Wagner, Z.~Harman, G.~Werth, W.~Quint,
  C.~H. Keitel, and K.~Blaum.
\newblock High-precision measurement of the atomic mass of the electron.
\newblock {\em Nature}, 506:467--470, 02 2014.

\bibitem{Moo14}
A.~Mooser, S.~Ulmer, K.~Blaum, K.~Franke, H.~Kracke, C.~Leiteritz, W.~Quint,
  C.~C. Rodegheri, C.~Smorra, and J.~Walz.
\newblock Direct high-precision measurement of the magnetic moment of the
  proton.
\newblock {\em Nature}, 509(7502):596--599, 05 2014.

\bibitem{Nag17}
H.~Nagahama, C.~Smorra, S.~Sellner, J.~Harrington, T.~Higuchi, M.~J. Borchert,
  T.~Tanaka, M.~Besirli, A.~Mooser, G.~Schneider, and et~al.
\newblock Sixfold improved single particle measurement of the magnetic moment
  of the antiproton.
\newblock {\em Nature Communications}, 8:14084 EP --, 01 2017.

\bibitem{Win90}
D.~J. Heinzen and D.~J. Wineland.
\newblock Quantum-limited cooling and detection of radio-frequency oscillations
  by laser-cooled ions.
\newblock {\em Phys. Rev. A}, 42:2977--2994, Sep 1990.

\bibitem{Fen96}
X.~Feng, M.~Charlton, M.~Holzscheiter, R.~A. Lewis, and Y.~Yamazaki.
\newblock Tank circuit model applied to particles in a penning trap.
\newblock {\em Journal of Applied Physics}, 79(1):8--13, 2017/06/18 1996.

\bibitem{Nag16}
H.~Nagahama, G.~Schneider, A.~Mooser, C.~Smorra, S.~Sellner, J.~Harrington,
  T.~Higuchi, M.~Borchert, T.~Tanaka, M.~Besirli, and et~al.
\newblock Highly sensitive superconducting circuits at ∼700 khz with tunable
  quality factors for image-current detection of single trapped antiprotons.
\newblock {\em Review of Scientific Instruments}, 87(11):113305, 2017/06/18
  2016.

\bibitem{Bro87}
L.~S. Brown and G.~Gabrielse.
\newblock Geonium theory: Physics of a single electron or ion in a penning
  trap.
\newblock {\em Rev. Mod. Phys.}, 58:233--311, Jan 1986.

\bibitem{Bro84}
L.~S. Brown.
\newblock Line shape for a precise measurement of the electron's magnetic
  moment.
\newblock {\em Phys. Rev. Lett.}, 52:2013--2015, Jun 1984.

\bibitem{Bro85}
L.~S. Brown.
\newblock Geonium lineshape.
\newblock {\em Annals of Physics}, 159(1):62--98, 1985.

\bibitem{Ulm11}
S.~Ulmer, C.~C. Rodegheri, K.~Blaum, H.~Kracke, A.~Mooser, W.~Quint, and
  J.~Walz.
\newblock Observation of spin flips with a single trapped proton.
\newblock {\em Phys. Rev. Lett.}, 106:253001, Jun 2011.

\bibitem{Moo13}
A.~Mooser, H.~Kracke, K.~Blaum, S.~A. Br\"auninger, K.~Franke, C.~Leiteritz,
  W.~Quint, C.~C. Rodegheri, S.~Ulmer, and J.~Walz.
\newblock Resolution of single spin flips of a single proton.
\newblock {\em Phys. Rev. Lett.}, 110:140405, Apr 2013.

\bibitem{Smo17}
C.~Smorra, A.~Mooser, M.~Besirli, M.~Bohman, M.~J. Borchert, J.~Harrington,
  T.~Higuchi, H.~Nagahama, G.~L. Schneider, S.~Sellner, and et~al.
\newblock Observation of individual spin quantum transitions of a single
  antiproton.
\newblock {\em Physics Letters B}, 769:1--6, 6 2017.

\bibitem{Smo15}
C.~Smorra, K.~Blaum, L.~Bojtar, M.~Borchert, K.~A. Franke, T.~Higuchi,
  N.~Leefer, H.~Nagahama, Y.~Matsuda, A.~Mooser, M.~Niemann, C.~Ospelkaus,
  W.~Quint, G.~Schneider, S.~Sellner, T.~Tanaka, S.~Van~Gorp, J.~Walz,
  Y.~Yamazaki, and S.~Ulmer.
\newblock Base --the baryon antibaryon symmetry experiment.
\newblock {\em The European Physical Journal Special Topics},
  224(16):3055--3108, 2015.

\bibitem{Sav97}
T.~A. Savard, K.~M. O'Hara, and J.~E. Thomas.
\newblock Laser-noise-induced heating in far-off resonance optical traps.
\newblock {\em Phys. Rev. A}, 56:R1095--R1098, Aug 1997.

\bibitem{Kos11}
V.~A. Kosteleck\'y and N.~Russell.
\newblock Data tables for lorentz and $cpt$ violation.
\newblock {\em Rev. Mod. Phys.}, 83:11--31, Mar 2011.

\bibitem{Stu11}
S.~Sturm, A.~Wagner, B.~Schabinger, and K.~Blaum.
\newblock Phase-sensitive cyclotron frequency measurements at ultralow
  energies.
\newblock {\em Phys. Rev. Lett.}, 107:143003, Sep 2011.

\bibitem{Mon16}
K.~R. Brown, J.~Kim, and C.~Monroe.
\newblock Co-designing a scalable quantum computer with trapped atomic ions.
\newblock {\em npj Quantum Information}, 2:16034, 11 2016.

\bibitem{Win95}
C.~Monroe, D.~M. Meekhof, B.~E. King, W.~M. Itano, and D.~J. Wineland.
\newblock Demonstration of a fundamental quantum logic gate.
\newblock {\em Phys. Rev. Lett.}, 75:4714--4717, Dec 1995.

\bibitem{Bla16}
E.~A. Martinez, C.~A. Muschik, P.~Schindler, D.~Nigg, A.~Erhard, M.~Heyl,
  P.~Hauke, M.~Dalmonte, T.~Monz, P.~Zoller, and et~al.
\newblock Real-time dynamics of lattice gauge theories with a few-qubit quantum
  computer.
\newblock {\em Nature}, 534(7608):516--519, 06 2016.

\bibitem{Mon95}
C.~Monroe, D.~M. Meekhof, B.~E. King, S.~R. Jefferts, W.~M. Itano, D.~J.
  Wineland, and P.~Gould.
\newblock Resolved-sideband raman cooling of a bound atom to the 3d zero-point
  energy.
\newblock {\em Phys. Rev. Lett.}, 75:4011--4014, Nov 1995.

\bibitem{Goo16}
J.~F. Goodwin, G.~Stutter, R.~C. Thompson, and D.~M. Segal.
\newblock Resolved-sideband laser cooling in a penning trap.
\newblock {\em Phys. Rev. Lett.}, 116:143002, Apr 2016.

\bibitem{Lar86}
D.~J. Larson, J.~C. Bergquist, J.~J. Bollinger, Wayne~M. Itano, and D.~J.
  Wineland.
\newblock Sympathetic cooling of trapped ions: A laser-cooled two-species
  nonneutral ion plasma.
\newblock {\em Phys. Rev. Lett.}, 57:70--73, Jul 1986.

\bibitem{Bli02}
B.~B. Blinov, L.~Deslauriers, P.~Lee, M.~J. Madsen, R.~Miller, and C.~Monroe.
\newblock Sympathetic cooling of trapped ${\mathrm{cd}}^{+}$ isotopes.
\newblock {\em Phys. Rev. A}, 65:040304, Apr 2002.

\bibitem{Bar03}
M.~D. Barrett, B.~DeMarco, T.~Schaetz, V.~Meyer, D.~Leibfried, J.~Britton,
  J.~Chiaverini, W.~M. Itano, B.~Jelenkovi\ifmmode~\acute{c}\else \'{c}\fi{},
  J.~D. Jost, C.~Langer, T.~Rosenband, and D.~J. Wineland.
\newblock Sympathetic cooling of ${}^{9}{\mathrm{be}}^{+}$ and
  ${}^{24}{\mathrm{mg}}^{+}$ for quantum logic.
\newblock {\em Phys. Rev. A}, 68:042302, Oct 2003.

\bibitem{Sch15}
L.~Schm{\"o}ger, M.~Schwarz, T.~M. Baumann, O.~O. Versolato, B.~Piest,
  T.~Pfeifer, J.~Ullrich, P.~O. Schmidt, and J.~R. Crespo L{\'o}pez-Urrutia.
\newblock Deceleration, precooling, and multi-pass stopping of highly charged
  ions in be+ coulomb crystals.
\newblock {\em Review of Scientific Instruments}, 86(10):103111, 2017/06/22
  2015.

\bibitem{Bro11}
K.~R. Brown, C.~Ospelkaus, Y.~Colombe, A.~C. Wilson, D.~Leibfried, and D.~J.
  Wineland.
\newblock Coupled quantized mechanical oscillators.
\newblock {\em Nature}, 471(7337):196--199, 03 2011.

\bibitem{Har11}
M.~Harlander, R.~Lechner, M.~Brownnutt, R.~Blatt, and W.~Hansel.
\newblock Trapped-ion antennae for the transmission of quantum information.
\newblock {\em Nature}, 471(7337):200--203, 03 2011.

\bibitem{Win75}
D.~J. Wineland and H.~G. Dehmelt.
\newblock Principles of the stored ion calorimeter.
\newblock {\em Journal of Applied Physics}, 46(2):919--930, 2017/06/18 1975.

\bibitem{Dan09}
N.~Daniilidis, T~Lee, R.~Clark, S.~Narayanan, and H.~H{\"a}ffner.
\newblock Wiring up trapped ions to study aspects of quantum information.
\newblock {\em Journal of Physics B: Atomic, Molecular and Optical Physics},
  42(15):154012, 2009.

\bibitem{Col14}
Y.~Colombe, D.~H. Slichter, A.~C. Wilson, D.~L., and David~J. Wineland.
\newblock Single-mode optical fiber for high-power, low-loss uv transmission.
\newblock {\em Optics Express}, 22(16):19783--19793, 2014.

\bibitem{nie14}
M.~Niemann, A.G. Paschke, T.~Dubielzig, S.~Ulmer, and C.~Ospelkaus.
\newblock Cpt test with (anti) proton magnetic moments based on quantum logic
  cooling and readout.
\newblock In {\em CPT and Lorentz Symmetry-Proceedings of the Sixth Meeting},
  volume~1, pages 41--44, 2014.

\end{thebibliography}

\end{document}